\documentstyle[aps,prl,preprint,floats,epsf]{revtex}

\input epsfig.sty

\newcommand{\ensuremath}[1]{\relax\ifmmode{#1}\else{$#1$}\fi}
\newcommand{\BDsDs}   {\ensuremath{\BZ\!\to\!\DsP\,\DsM}}
\newcommand{\BDSDS}   {\ensuremath{\BZ\!\to\!\DSP\,\DSM}}
\newcommand{\BDssDs}  {\ensuremath{\BZ\!\to\!\DssP\,\DsM}}
\newcommand{\BDSD}    {\ensuremath{\BZ\!\to\!\DSD}}
\newcommand{\BDD}     {\ensuremath{\BZ\!\to\!\DP\,\DM}}
\newcommand{\BPSIKS}  {\ensuremath{\BZ\!\to\!\PSI\,\KS}}
\newcommand{\DSPIDZ}  {\ensuremath{\DSP\!\to\!\PIP\,\DZ}}
\newcommand{\DZKPI}   {\ensuremath{\DZ\!\to\!\KM\,\PIP}}
\newcommand{\DZKPIPIZ}{\ensuremath{\DZ\!\to\!\KM\,\PIP\,\PIZ}}
\newcommand{\DZKPPP}  {\ensuremath{\DZ\!\to\!\KM\,\PIP\,\PIM\,\PIP}}
\newcommand{\DPKPIPI} {\ensuremath{\DP\!\to\!\KM\,\PIP}\,\PIP}

\newcommand{\B}   {\ensuremath{B}}
\newcommand{\BB}  {\ensuremath{\bar{B}}}
\newcommand{\BZ}  {\ensuremath{B^0}}

\newcommand{\DS}  {\ensuremath{D^*}}
\newcommand{\DSP} {\ensuremath{D^{*+}}}
\newcommand{\DSM} {\ensuremath{D^{*-}}}
\newcommand{\Ds}  {\ensuremath{D^{(*)}}}
\newcommand{\DsP} {\ensuremath{D^{(*)+}}}
\newcommand{\DsM} {\ensuremath{D^{(*)-}}}
\newcommand{\D}   {\ensuremath{D}}
\newcommand{\DP}  {\ensuremath{D^+}}
\newcommand{\DM}  {\ensuremath{D^-}}
\newcommand{\DZ}  {\ensuremath{D^0}}

\newcommand{\DSD} {\ensuremath{D^{*\pm}\,D^\mp}}

\newcommand{\DssP}{\ensuremath{D^{(*)+}_{s}}}

\newcommand{\PSI} {\ensuremath{\psi}}

\newcommand{\KM}  {\ensuremath{K^-}}
\newcommand{\KS}  {\ensuremath{K^0_S}}

\newcommand{\PIP} {\ensuremath{\pi^+}}
\newcommand{\PIM} {\ensuremath{\pi^-}}

\newcommand{\PIZ} {\ensuremath{\pi^0}}

\newcommand{\COSHEL}{\ensuremath{\cos\theta_{\pi^+}}}
\newcommand{\DELE}  {\ensuremath{\Delta E}}
\newcommand{\DELM}  {\ensuremath{\Delta m_{\DS-\D}}}
\newcommand{\BCM}   {\ensuremath{m_B}}

\newcommand{\CHDSDSa} {\ensuremath{
           \left(\frac{({\Delta m})_1 - \langle{\Delta m}\rangle}
                      {\sigma_{\Delta m}}\right)^2 +
           \left(\frac{({\Delta m})_2 - \langle{\Delta m}\rangle}
                      {\sigma_{\Delta m}}\right)^2 + }}
\newcommand{\CHDSDSb} {\ensuremath{
           \left(\frac{(m_{\DZ})_1 - \langle m_{\DZ}\rangle}
                      {\sigma_{m_{\DZ}}}\right)^2 +
           \left(\frac{(m_{\DZ})_2 - \langle m_{\DZ}\rangle}
                      {\sigma_{m_{\DZ}}}\right)^2  }}
\newcommand{\CHDSD} {\ensuremath{
           \left(\frac{{\Delta m} - \langle{\Delta m}\rangle}
                      {\sigma_{\Delta m}}\right)^2 +
           \left(\frac{m_{\DZ} - \langle m_{\DZ}\rangle}
                      {\sigma_{m_{\DZ}}}\right)^2 +
           \left(\frac{m_{\DP} - \langle m_{\DP}\rangle}
                      {\sigma_{m_{\DP}}}\right)^2  }}
\newcommand{\CHDD} {\ensuremath{
           \left(\frac{(m_{\DP})_1 - \langle m_{\DP}\rangle}
                      {\sigma_{m_{\DP}}}\right)^2 +
           \left(\frac{(m_{\DP})_2 - \langle m_{\DP}\rangle}
                      {\sigma_{m_{\DP}}}\right)^2  }}
\newcommand{\BRDSDS}{\ensuremath{(5.3^{+7.1}_{-3.7}(\rm{stat})\pm 
                     1.0(\rm{syst}))\times 10^{-4}}}
\newcommand{\ULDSDS}{\ensuremath{2.2 \times 10^{-3}}}
\newcommand{\ULDSD} {\ensuremath{1.8 \times 10^{-3}}}
\newcommand{\ULDD}  {\ensuremath{1.2 \times 10^{-3}}}
\newcommand{\BGDSDS}{\ensuremath{0.022 \pm 0.011}}

\begin{document}
\preprint{\tighten\vbox{\hbox{\hfil CLNS 97/1474}
                        \hbox{\hfil CLEO 97-6}
}}


\title{Search for the Decays \BDsDs}  

\author{CLEO Collaboration}
\date{\today}

\maketitle
\tighten

\begin{abstract} 
Using the CLEO-II data set
we have searched for the Cabibbo-suppressed decays {\BDsDs}.
For the decay {\BDSDS}, we observe one candidate signal event,
with an expected
background of {\BGDSDS} events.
This yield corresponds to a branching fraction 
of ${\cal B}(\BDSDS) = {\BRDSDS}$
and an upper limit of ${\cal B}(\BDSDS) < \ULDSDS$ at
the $90\%$ CL.
For {\BDSD} and {\BDD},
no significant excess of signal above the expected background level is seen,
and we calculate the $90\%$ CL upper limits on the branching fractions to 
be ${\cal B}(\BDSD) < \ULDSD$ and ${\cal B}(\BDD) < \ULDD$.
\end{abstract}


{
\renewcommand{\thefootnote}{\fnsymbol{footnote}}

\begin{center}
D.~M.~Asner,$^{1}$ D.~W.~Bliss,$^{1}$ W.~S.~Brower,$^{1}$
G.~Masek,$^{1}$ H.~P.~Paar,$^{1}$ V.~Sharma,$^{1}$
J.~Gronberg,$^{2}$ R.~Kutschke,$^{2}$ D.~J.~Lange,$^{2}$
S.~Menary,$^{2}$
% ...
\end{center}
 
\small
\begin{center}
$^{1}${University of California, San Diego, La Jolla, California 92093}\\
$^{2}${University of California, Santa Barbara, California 93106}\\
% ...
$^{13}${University of Minnesota, Minneapolis, Minnesota 55455}\\
$^{14}${State University of New York at Albany, Albany, New York 12222}\\
% ...
$^{25}${California Institute of Technology, Pasadena, California 91125}
\end{center}
\setcounter{footnote}{0}
}
\newpage

The decays {\BDsDs}
are favorable modes for studying $CP$ violation in $B$ decays.
In the Standard Model, time-dependent asymmetries in the decays 
can be related to the angle $\beta$ of the unitarity triangle \cite{Aleksan}.
This angle can also be measured with
{\BPSIKS} decays;
any difference between the
values obtained in {\BDsDs} decays and {\BPSIKS} would indicate
non-Standard Model mechanisms for $CP$ violation
\cite{BaBarTDR,CLEO-B}.
Although {\BDSDS} and {\BDSD} are not pure $CP$
eigenstates, estimates indicate that a dilution of
the $CP$ asymmetry
of only a few percent would be incurred by treating these modes as pure $CP$
eigenstates \cite{Aleksan}.

The modes {\BDsDs} have never been observed, and no published limits
on their branching fractions exist.
The decay amplitude is dominated by a spectator diagram with
$\bar{b} \to \bar{c}W^+$ followed by the Cabibbo-suppressed
process $W^+ \to c\bar{d}$.
One can estimate the branching fractions
for {\BDsDs} by relating them to the Cabibbo-favored decays {\BDssDs}:
\begin{eqnarray}
{\cal B}(\BDsDs) & \simeq &
\left(\frac{f_{D^{(*)}}}{f_{D_s^{(*)}}}\right)^2 \tan^2\theta_C \ 
{\cal B}(\BDssDs),
\end{eqnarray}
where the $f_X$ are decay constants and $\theta_C$ is the Cabibbo angle.
Table~\ref{tab:EstBr} shows the expected {\BDsDs} branching fractions, where
the CLEO measurements of ${\cal B}(\BDssDs)$ have been used \cite{CLEODsD}. 

\begin{table}[htb]
\begin{center}
\caption{Estimated branching fractions for {\BDsDs} based on the measured
branching fractions of the Cabibbo-favored decays {\BDssDs}.}
\label{tab:EstBr}
\begin{tabular}{lcc}
Mode & ${\cal B}$ of Related & Estimated ${\cal B}$ for \\
     &  {\DssP}{\DsM} Mode (\%)      &  {\DsP}{\DsM} ($10^{-4}$)\\ \hline
\BDSDS & 2.4 & 9.7 \\
\BDSD  & 2.0 & 8.1 \\
\BDD   & 1.1 & 4.5 \\
\end{tabular}
\end{center}
\end{table}

The data used in this analysis were recorded with the CLEO-II detector
\cite{CLEO-II} located at the Cornell Electron Storage Ring (CESR).
An integrated luminosity of $3.09~{\rm fb}^{-1}$ was taken at the
$\Upsilon (4{\rm S})$ resonance, corresponding to approximately
$3.3 \times 10^6~{\B\BB}$ pairs produced.

At the $\Upsilon(4{\rm S})$, the {\B\BB} pairs are produced nearly at rest,
resulting in a spherical event topology.  In contrast,
non-{\B\BB},
continuum events have a more jet-like topology.  To select spherical
events we required that the ratio $R_2$ of the second and zeroth Fox-Wolfram
moments \cite{FoxWolfram} be less than 0.25.

We required charged tracks to be of good quality
and consistent with coming
from the interaction point in both the $r-\phi$ and $r-z$ planes.  We defined
photon candidates as isolated clusters in the CsI calorimeter with energy
greater than 30~MeV in the central region ($\cos\theta\leq 0.71$,
where $\theta$ is measured from the beamline) and
greater than 50~MeV elsewhere.  Pairs of photons with measured invariant
masses within 2.5 standard deviations of the nominal \PIZ\
mass were used to form \PIZ\ candidates.
Selected \PIZ\ candidates were then kinematically fitted to the
nominal \PIZ\ mass.

A particle identification system consisting of $dE/dx$ and
time-of-flight was used to distinguish charged kaons from charged pions.
For charged pion candidates, we required the likelihood of the pion
hypothesis, $L_{\pi}$, to be greater than 0.05.
Since all signal modes require
two charged kaons, the kaon candidates were required to have a joint
kaon hypothesis likelihood, $L_{K_1} L_{K_2}$, greater than 0.10.

We reconstructed all {\DSP} candidates in the mode \DSPIDZ\
(charge-conjugate modes are implied).  \DZ\ candidates were reconstructed
in the modes {\DZKPI}, {\DZKPIPIZ} and {\DZKPPP}.  \DP\ candidates were
reconstructed via {\DPKPIPI}.  Table~\ref{tab:Modes} summarizes the
branching fractions of the {\Ds} modes used \cite{PDG}.

\begin{table}[htb]
\caption{Branching fractions of {\Ds} modes used in reconstruction.}
\label{tab:Modes}
\begin{center}
\begin{tabular}{lc}
Decay Mode   & Branching Fraction (\%)\\ \hline 
\DSPIDZ      & $68.3 \pm 1.4$ \\ 
\DZKPI       & $3.83 \pm 0.12$ \\
\DZKPIPIZ    & $13.9 \pm 0.9$ \\
\DZKPPP      & $7.5 \pm 0.4$  \\ 
\DPKPIPI     & $9.1 \pm 0.6$  \\
\end{tabular}
\end{center}
\end{table}

For the decay mode {\DZKPIPIZ}, we make a cut on the weight in the
Dalitz plot in order to take advantage of the resonant substructure
present in the decay.
The cut choosen was 76\% efficient for good {\DZKPIPIZ} decays while
rejecting 69\% of the background.

We performed a vertex-constrained fit on all the charged tracks in the
\BZ\ candidate for modes that contained a {\DSP}.
The $\chi^2$ from the vertex fit
was required to be less than 100.  The fit
improved the determination of the angular track parameters for the
slow \PIP\ from the \DSP\ decay.  The resulting r.\@m.\@s.\ resolution on
the reconstructed mass difference
$\DELM \equiv m_{\DSP} - m_{\DZ}$ was approximately 0.69~MeV.

Because \BDSD\ is a $Pseudoscalar \to Vector + Pseudoscalar$ decay, 
the cosine of the decay angle,
{\COSHEL}, of the slow {\PIP} from the
{\DSP} has a
$\cos^2\theta$ distribution, while
background events have a uniform distribution
in this variable.  For \BDSD\ candidates we required $|\COSHEL| > 0.5$.

To select {\BZ} candidates that contain well-identified {\Ds}s we
combine the reconstructed {\Ds} masses into a single quantity, $\chi^2_M$. 
The definition of $ \chi^2_M $ for each mode is given by
\begin{eqnarray}
\nonumber \chi^2_M ({\DSP\DSM}) & = & {\CHDSDSa} \\
                      &   & {\CHDSDSb} \\
\chi^2_M ({\DSD})     & = & {\CHDSD}  \\
\chi^2_M ({\DP\DM})   & = & {\CHDD},
\end{eqnarray}
where the values in angle brackets represent the nominal values and the
sigmas are the r.\@m.\@s.\ resolutions on the given quantity.
We require
    $\chi^2_M ({\DSP\DSM}) < 8.0$,
    $\chi^2_M ({\DSD}) < 4.0$
and $\chi^2_M ({\DP\DM}) < 2.0$.
From studies of Monte Carlo and regions
in the data outside of the signal areas in other variables, we
find that the backgrounds are uniform in
$\chi^2_M$.

Since the energy of the \BZ\ is equal to the beam energy at CESR,
we used the beam
energy instead of the measured energy of the {\BZ} candidate to calculate
the beam-constrained mass: $\BCM = \sqrt{E_{beam}^2 - {\bf p}_B^2}$.
The r.\@m.\@s.\ resolution in {\BCM} for signal events,
as determined from Monte Carlo,
is $2.8$~MeV.  In addition, the energy difference, $\DELE \equiv
E_{B} - E_{beam}$, where $E_B$ is the measured \BZ\ energy,
was used to distinguish signal from background.  The
resolution in {\DELE} is 12~MeV after performing a mass-constrained fit
that included the masses of all secondary particles ({\Ds} and {\PIZ}).
The signal region in all modes was defined as $|\DELE| < 2 \sigma_{\DELE}$
and $|\BCM - \langle{m_{B^0}}\rangle| < 2 \sigma_{\BCM}$.

We used a Monte Carlo simulation of the CLEO-II detector to optimize
all cuts.
Since the number
of observed signal events was expected to be small, all cuts were optimized
to minimize the probability that the expected background level 
would fluctuate
up to or beyond the expected signal level.  For
calculating the expected number of signal events during this optimization
we assumed a
branching fraction of 0.1\% for all \BDsDs\ modes.

Using the cuts defined above, we determined the signal reconstruction  
efficiency using
Monte Carlo.  The reconstruction  efficiency and single event sensitivity
($SES \equiv (\epsilon \ {\cal B} \ N_{\B\BB})^{-1}$, where
$\epsilon$ is the detection efficiency, ${\cal B}$ is the product of the
daughter branching fractions and $N_{\B\BB}$ is the number of {\B\BB} pairs
produced in the data set) for
each mode are summarized in Table~\ref{tab:Eff}.
The systematic uncertainty on the $SES$ is dominated largely by uncertainties
in the \D\ and \DS\ branching fractions, and, due to the large
mean multiplicity
of the final states, the uncertainty in the tracking efficiency.
\begin{table}[htb]
\begin{center}
\caption{Summary of reconstruction
         efficiencies and single
         event sensitivities for the three
         \BDsDs\ modes.}
\label{tab:Eff}
\begin{tabular}{lccc}
Mode   & Efficiency, $\epsilon$ & $SES \equiv (\epsilon \ {\cal B}
                                  \ N_{\B\BB})^{-1}$ \\
       & (\%)                   &  ($10^{-4}$)\\ \hline
\BDSDS & 1.86     & $5.45\pm 0.99$ \\
\BDSD  & 5.07     & $3.79\pm 0.53$ \\
\BDD   & 14.41    & $2.52\pm 0.40$ \\
\end{tabular}
\end{center}
\end{table}

The dominant background is due to random combinations from \B\BB\ and
continuum events.  The Monte Carlo predicts that this background varies
smoothly in \DELE\ and \BCM\ , and this is verified in the data.
The \BCM\ distribution for data
in \DELE\ sidebands ($50~{\rm MeV} \leq |\DELE| \leq 400~{\rm MeV}$)
varies smoothly with no peaking in the signal region.  The same is true for the
\DELE\ distribution for data with $\BCM < 5.27~{\rm GeV}$.
To estimate the background in the signal region, we count the events
in a sideband in the {\DELE}-\BCM\ plane
($50~{\rm MeV} \leq |\DELE| \leq 400~{\rm MeV}$;
$5.2~{\rm GeV} \leq \BCM \leq E_{beam}$) and multiply
by the relative efficiencies of the signal and sideband regions determined
from background Monte Carlo.

\begin{figure}[htb]
\begin{center}
\mbox{\psfig{file=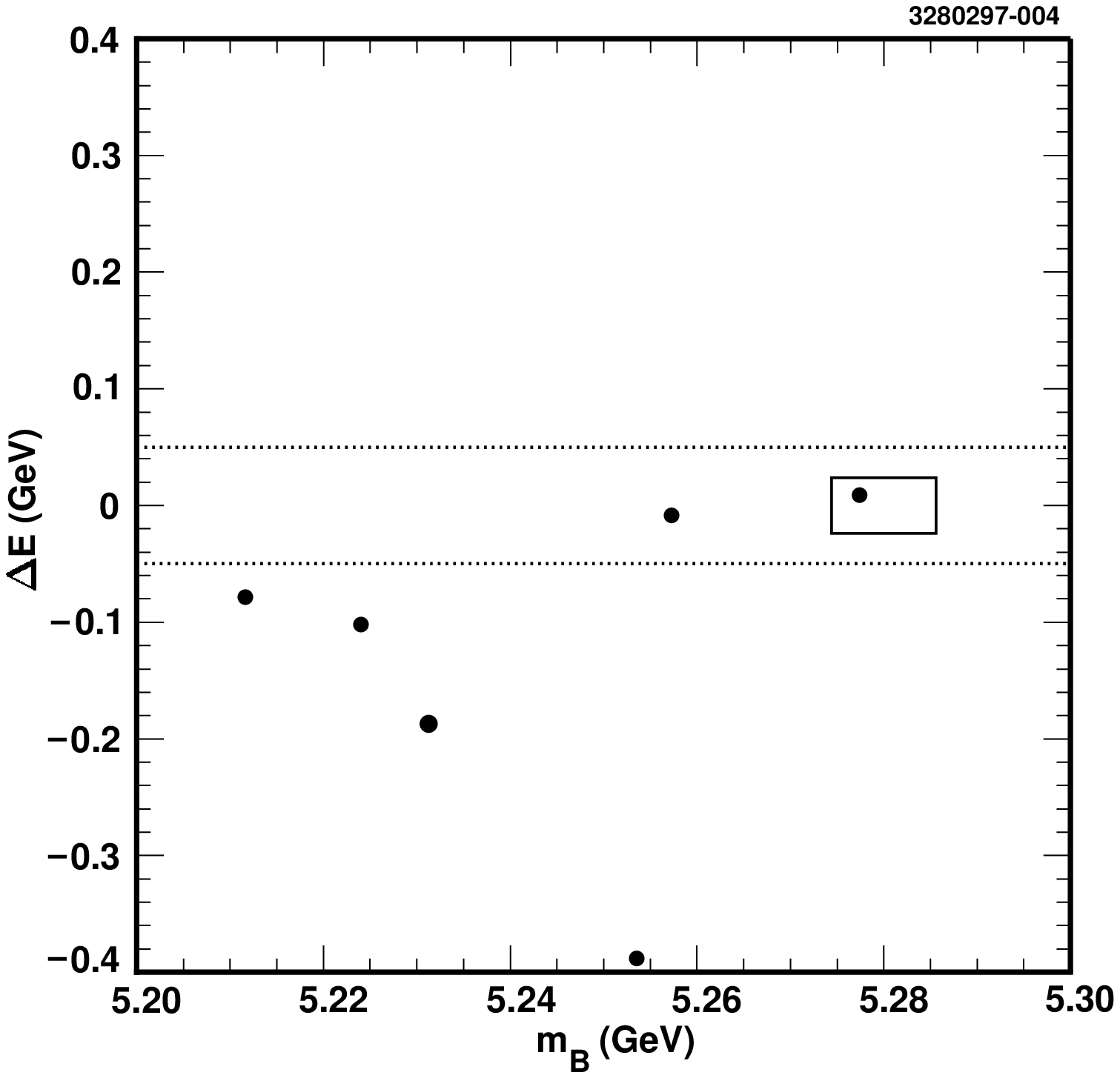,width=\textwidth,angle=0}}
\caption{\DELE\ vs.~\BCM\ for data in the \BDSDS analysis.  The signal region
         is indicated by a solid box.  The sideband region lies
         above the top and below the bottom dotted lines.}
\label{fig:DSDS}
\end{center}
\end{figure}

\begin{figure}[htb]
\begin{center}
\mbox{\psfig{file=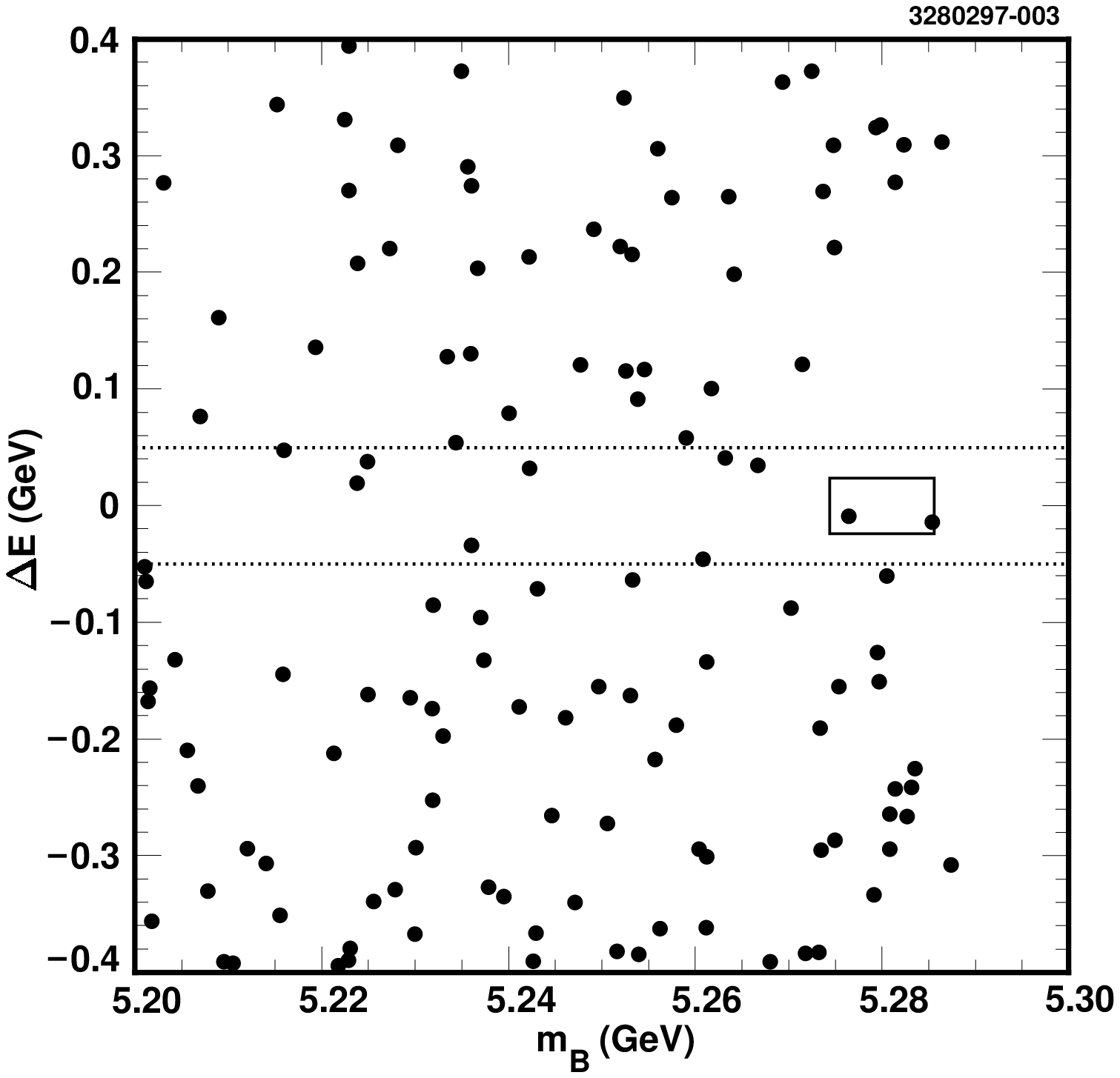,width=\textwidth,angle=0}}
\caption{\DELE\ vs.~\BCM\ for data in the \BDSD analysis.  The signal region
         is indicated by a solid box.  The sideband region lies
         above the top and below the bottom dotted lines.}
\label{fig:DSD}
\end{center}
\end{figure}

\begin{figure}[htb]
\begin{center}
\mbox{\psfig{file=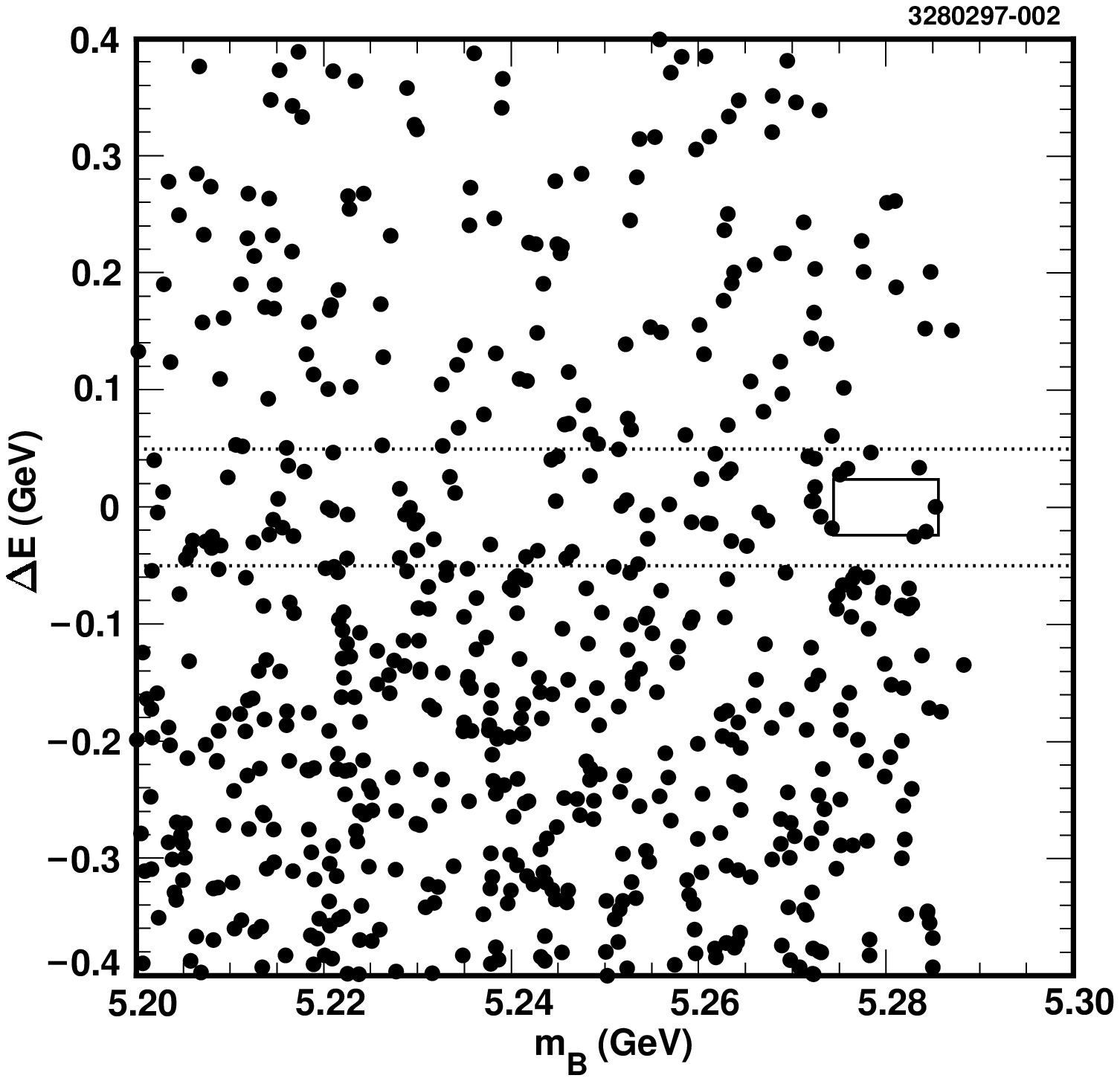,width=\textwidth,angle=0}}
\caption{\DELE\ vs.~\BCM\ for data in the \BDD analysis.  The signal region
         is indicated by a solid box.  The sideband region lies
         above the top and below the bottom dotted lines.}
\label{fig:DD}
\end{center}
\end{figure}

Figures~\ref{fig:DSDS}, \ref{fig:DSD} and \ref{fig:DD} show the resulting
plots of \DELE\ vs.~\BCM\ for the three modes.  The signal region
is indicated with a solid line, and the sideband region is indicated with
a dotted line.

Table~\ref{tab:Dat} lists the event yields
in the sideband and signal regions.
The expected number of background events in the signal region is also given.
The uncertainty on the expected number of
background events is a combination of statistical error on the number of
events in the \DELE\ sideband regions and the uncertainty in the background
shape through the signal region.
\begin{table}[htb]
\begin{center}
\caption{Summary of events found in the data, both in the \DELE\ sidebands
         and in the signal region, for each of the three modes.}
\label{tab:Dat}
\begin{tabular}{lccc}
Mode    & Events in \DELE\ & Predicted Background & Events found in \\
        & Sidebands        & in the Signal Region & Signal Region   \\ \hline
\BDSDS  & 4                & \BGDSDS              & 1               \\
\BDSD   & 117              & $0.64 \pm 0.10$      & 2               \\
\BDD    & 539              & $2.64 \pm 0.34$      & 3               \\
\end{tabular}
\end{center}
\end{table}

The probability that the expected background of \BGDSDS\ events in \BDSDS\
fluctuates up to one or more events is 2.2\%.
If we interpret the one observed  event as evidence for a signal,
the resulting branching fraction would be
\begin{equation}
{\cal B}(\BDSDS) = \BRDSDS ,
\end{equation}
where the systematic uncertainty comes from the uncertainty in the $SES$.

No significant excess of events is seen in the other two modes.  We
calculate upper limits on the branching fractions for all three modes, and
these results are summarized in Table~\ref{tab:Upper}.  The systematic
uncertainty in the $SES$ and the uncertainty in the background level
have been incorporated into the upper limits \cite{Cousins}.

\begin{table}[htb]
\begin{center}
\caption{Summary of upper limits on the \BDsDs\ branching fractions.  All
         upper limits are quoted at the 90\% confidence level.}
\label{tab:Upper}
\begin{tabular}{lc}
Mode   &   Upper Limit (90\% CL)  \\ \hline
\BDSDS &   \ULDSDS \\
\BDSD  &   \ULDSD  \\
\BDD   &   \ULDD   \\
\end{tabular}
\end{center}
\end{table}

We have performed a search for the decays {\BDsDs}.  In the mode {\BDSDS},
one event is seen in the signal region where the expected background is
{\BGDSDS}.
The one event in {\BDSDS} is seen at a rate
that is consistent with predictions, and in all three modes the upper
limits are within about a factor of two from the predicted branching fractions.

\begin{acknowledgments}
We gratefully acknowledge the effort of the CESR staff in providing us with
excellent luminosity and running conditions.
This work was supported by 
the National Science Foundation,
the U.S. Department of Energy,
the Heisenberg Foundation,  
the Alexander von Humboldt Stiftung,
%
Research Corporation,
the Natural Sciences and Engineering Research Council of Canada,
and the A.P. Sloan Foundation.

\end{acknowledgments}


\begin{thebibliography}{99}

\bibitem{Aleksan}
	R.~Aleksan {\it et al.}, Phys.~Lett.~{\bf B317}, 173 (1993).
\bibitem{BaBarTDR}
	The {\it BaBar} Collaboration, Technical Design Report, SLAC-R-95-457
	(1995). 
\bibitem{CLEO-B}
	K.~Lingel {\it et al.}, ``Physics Rationale for a B Factory'',
	CLNS 91-1043 (1991).
\bibitem{CLEODsD}
	CLEO Collaboration, D.~Gibaut {\it et al.}, Phys.~Rev.~D {\bf 53},
	4734 (1996). 
\bibitem{CLEO-II}
	CLEO Collaboration, Y.~Kubota {\it et al.}, Nucl.~Instrum.~Methods A
	{\bf 320}, 66 (1992).
\bibitem{FoxWolfram}
	G.~Fox and S.~Wolfram, Phys.~Rev.~Lett.~{\bf 41}, 1581 (1978).
\bibitem{PDG}
        Particle Data Group, R.~M.~Barnett {\it et al.}, Phys.~Rev.~D {\bf 54},
        1 (1996).
\bibitem{Cousins}
	R.~D.~Cousins and V.~Highland, Nucl.~Instrum.~Methods A {\bf 320},
        331 (1992).

\end{thebibliography}
\end{document}